\documentclass[aps, pra, twocolumn, showpacs, superscriptaddress]{revtex4-1}

\usepackage{amsmath,amssymb}
\usepackage{graphicx}
\usepackage{dcolumn}
\usepackage{bm}
\usepackage{color,xcolor}
\begin{document}

\title {Simulating heavy fermion physics in optical lattice: Periodic Anderson model with harmonic trapping potential}
\author{Yin Zhong}
\email{zhongy05@hotmail.com}
\affiliation{Center for Interdisciplinary Studies $\&$ Key Laboratory for
Magnetism and Magnetic Materials of the MoE, Lanzhou University, Lanzhou 730000, China}

\author{Yu Liu}
\affiliation{LCP, Institute of Applied Physics and Computational Mathematics, Beijing 100088, China}
\affiliation{Software Center for High Performance Numerical Simulation,China Academy of Engineering Physics, Beijing 100088, China}

\author{Hong-Gang Luo}
\affiliation{Center for Interdisciplinary Studies $\&$ Key Laboratory for
Magnetism and Magnetic Materials of the MoE, Lanzhou University, Lanzhou 730000, China}
\affiliation{Beijing Computational Science Research Center, Beijing 100084, China}

\date{\today}

\begin{abstract}
Periodic Anderson model (PAM), where local electron orbitals interplay with itinerant electronic carriers, plays an essential role in our understanding on heavy fermion materials.
Motivated by recent proposal of simulating Kondo lattice model (KLM) in terms of alkaline-earth metal atoms, we make a further step toward simulation of PAM, which includes
crucial charge/valence fluctuation of local f-electron beyond purely low-energy spin fluctuation in KLM. To realize PAM, transition induced by suitable laser between electronic excited and ground state
of alkaline-earth metal atoms ($^{1}S_{0}$$\rightleftharpoons$$^{3}P_{0}$) is introduced, and it leads to effective hybridization between local electron and conduction electron in PAM.
Generally, the $SU(N)$ version of PAM can be realized by our proposal, which gives a unique opportunity to detect large-$N$ physics without complexity in realistic materials. In the present work, high temperature physical feature of standard ($SU(2)$) PAM with harmonic trapping potential is detailed analyzed by quantum Monte Carlo and dynamic mean-field theory. Indications for near-future experiments are provided. We expect our theoretical proposal and (hopefully) forthcoming experiments will deepen our understanding on heavy fermion systems and at the same time triggers further studies on related Mott physics, quantum criticality and non-trivial topology in both inhomogeneous and nonequilibrium realm.
\end{abstract}

\maketitle

\section{Introduction}
Understanding emergent novel collective quantum effects in heavy fermion system, e.g. quantum criticality, strange metal, hidden order and unconventional superconductivity\cite{Hewson1993,Coleman2015,Tsunetsugu1997,Rosch2007,Pfleiderer2009,Mydosh2011,Yang2016},
is one of most fundamental issues in modern condensed matter physics.
In these materials, electronic effective mass is gigantically enhanced compared to normal metals and local f-electron orbits interplay with itinerant charge carriers,
which is believed to be an essential driven force for those mentioned emergent phenomena\cite{Doniach1977}.
However, in spite of decades of intensive theoretical and experimental efforts, the nature of these systems is still poorly understood and even the most simplified model Hamiltonian like
Kondo lattice (KLM) and periodic Anderson lattice (PAM) defy our investigation due to notorious fermion minus-sign problem\cite{Vekic1995,Assaad1999}.

\begin{figure}
\includegraphics[width=0.7\columnwidth]{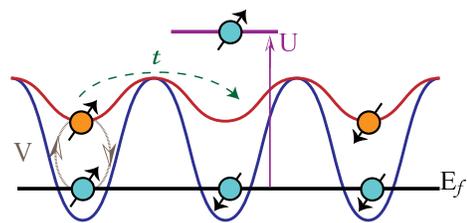}
\caption{\label{fig:PAM} Periodic Anderson model describes hybridization $V$ between local electron orbital (blue) and conducting charge carrier (orange). The single occupied local electron has energy $E_{f}$ while double occupation has extra Coloumb energy $U$. The conduction electron hops $t$ between nearest-neighbor sites and it gives a conducting energy band. For simplicity, external harmonic trapping potential is not shown here. }
\end{figure}
Fortunately, thanks to the rapid development of ultracold fermionic atom loaded in optical lattice\cite{Kohl2005,Jordens2008,Hart2015,Greif2016,Cheuk2016,Parsons2016,Boll2016}, 
intriguing plan to simulate Kondo lattice model in terms of alkaline-earth metal atoms $^{87}$Sr and $^{173}$Yb is recently proposed
and preliminary theoretical works in this direct have been achieved\cite{Gorshkov2010,Foss-Feig2010a,Foss-Feig2010b,Silva-Valencia2012,Jiang2014,Isaev2015,Isaev2016,ZhangRen2016}. More recently, interaction in these alkaline-earth atom gases has been under control in terms of the newly proposed orbital Feshbach resonance\cite{ZhangRen2015,Pagano2015,Riegger2015}. (Although $^{173}$Yb is in practice a rare-earth atom, it has similar electronic structure to alkaline-earth metal atom and we will only use the name 'alkaline-earth metal atoms' for simplicity.) In realistic heavy fermion compounds,
charge/valence fluctuation of local f-electron plays a crucial role in many mixed valence materials including the hotly studied candidate of \emph{topological} Kondo insulator SmB$_{6}$\cite{Dzero2016}.
Furthermore, valance fluctuation is also argued to be responsible for the second superconducting phases in prototypical heavy fermion superconductor CeCu$_{2}$Si$_{2}$ and
volume-changed transition in monovalent Ce under high pressure\cite{Yuan2003,Rueff2006}. More theoretically, the orbital-selective Mott transition proposed for heavy fermion quantum criticality is rooted on the localization of f-electron driven by large charge fluctuation, as well\cite{Pepin2007,Zhong2012}.

Intrigued by those interesting phenomena, the periodic Anderson model, (see also Fig.~\ref{fig:PAM}) which is a lattice extension of well-known single impurity Anderson model and is believed to capture such charge/valence fluctuation in the simplest fashion, should be taken into fully account though to our knowledge there are still no proposals to realize this fundamental model Hamiltonian in cold fermion systems.

\begin{figure}
\includegraphics[width=0.7\columnwidth]{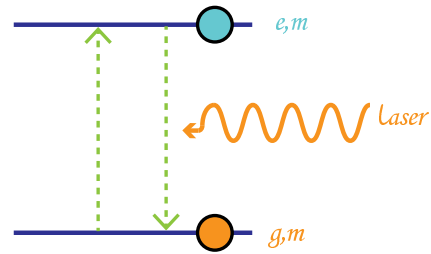}
\caption{\label{fig:scheme} Extra laser field is applied to induce direct transition between electronic excited ($e,m$) and ground ($g,m$) state, e.g. $^{1}S_{0}$$\rightleftharpoons$$^{3}P_{0}$ (nuclear Zeeman state denoted by $m$ is intact) which is able to simulate an effective hybridization $V$ between local electron and conduction electron in periodic Anderson model. (For simplicity, only coupling with the same Zeeman sub-level is considered here.)}
\end{figure}

In this article, we pave the first step to this issue with providing a simple scheme to realize periodic Anderson model in alkaline-earth metal atoms
and analyze basic features of this model with state of art numerical tool, the determinant quantum Monte Carlo (DQMC) and supplemented with dynamic mean-field theory (DMFT)\cite{Blankenbecler1981,Hirsch1985,Santo2003,Georges1996}, which correctly captures local quantum many-body dynamics and its prediction becomes exact at infinite dimension.
The main point of the scheme is that transition between electronic excited state $^{3}P_{0}$ and ground state $^{1}S_{0}$
of alkaline-earth metal atoms like $^{87}$Sr and $^{173}$Yb is driven by suitably chosen laser field, which is used to give an effective $c-f$ hybridization between local electrons and conduction ones in PAM. (See also Fig.~\ref{fig:scheme}.) Here, electronic excited state mimics the local f-electron and the ground state represents the conduction electron but without spin degree of freedom. Meanwhile, the nuclear Zeeman state, which is used to denote the effective/peusdo spin state in PAM, is intact. Combining these elements is able to provide a potential scheme to simulate PAM with proper tuning of experimental parameters
as proposed for KLM and have done in fermionic Hubbard model\cite{Kohl2005,Jordens2008,Gorshkov2010,Foss-Feig2010a}.
We should emphasize that in contrast to usual PAM in condensed matter physics, because external trapping potential always exists in optical lattice system,
PAM in cold atom simulation has to be supplemented with such (harmonic) trapping potential, whose simulation proposal and calculation are not reported in existing literature but are achieved in this work.

Interestingly, the general $SU(N)$ version of PAM is able to be realized in our proposal. Although the heavy fermion materials with unfilled f-shell usually have
such large degeneracy and the widely-used large-$N$ expansion is controllable in this artificial limit\cite{Coleman2015}, the complex chemical environment in realistic materials obscures the underlying physics
and the reliability of large-$N$ theory is questioned.
In contrast, the purity and tunability of optical lattice system gives us an unique opportunity to directly detect the large-N physics of PAM and inspect what has been captured by current theoretical understading.

Furthermore, since reachable temperature is comparatively high in present ultracold atom field, numerical simulations of DQMC and DMFT provide the basic physics at such high temperature regime, where Mott or orbital-selective Mott phenomena is clearly seen in the spatial dependent density distribution and energy-sensitive density of state. Importantly, these quantities are able to be measured by current cold atom techniques and future developments on alkaline-earth metal atoms are expected to realize
the proposed PAM and to detect its intriguing nature of quantum many-body interaction.
Moreover, if the standard PAM could be constructed,
the more exciting topological PAM involving subtle interplay between electron correlation and non-trivial band topology encoded by spin-orbital coupling
may be a next playground, where the idea of mentioned topological Kondo insulator can be explored without intrinsic complexity in realistic materials as SmB$_{6}$\cite{Dzero2016}.

The remainder of this paper is organized as follows.
In Sec.~\ref{sec2}, the proposal to realize PAM is detailed analyzed and PAM Hamiltonian is derived from general second quantized Hamiltonian for alkaline-earth metal atoms.
In Sec.~\ref{sec3}, high temperature properties are discussed and Mott physics of f-electron in harmonic trapping potential is explored. Finally, Sec.~\ref{sec4} is devoted to a brief
discussion and conclusion.
\section{Proposal to realize periodic Anderson model}\label{sec2}
The general second quantized Hamiltonian for alkaline-earth metal atoms reads,
\begin{eqnarray}
&&H=H_{G}+H_{hy}\nonumber \\
&&H_{G}=\sum_{\alpha m}\int d^{3}\mathbf{r} \Psi_{\alpha m}^{\dag}(\mathbf{r})(-\frac{\hbar^{2}}{2M}\nabla^{2}+V_{\alpha}(\mathbf{r})) \Psi_{\alpha m}(\mathbf{r})\nonumber \\
&&+\hbar\omega_{0}\int d^{3}\mathbf{r} (\rho_{e}(\mathbf{r})-\rho_{g}(\mathbf{r}))+\frac{g_{eg}^{+}+g_{eg}^{-}}{2}\int d^{3}\mathbf{r}\rho_{e}(\mathbf{r})\rho_{g}(\mathbf{r})\nonumber \\
&&+\sum_{\alpha,m<m'}g_{\alpha\alpha}\int d^{3}\mathbf{r}\rho_{\alpha m}(\mathbf{r})\rho_{\alpha m'}(\mathbf{r})\nonumber \\
&&+\frac{g_{eg}^{+}-g_{eg}^{-}}{2}\sum_{mm'}\int d^{3}\mathbf{r}\Psi_{g m}^{\dag}(\mathbf{r})\Psi_{e m'}^{\dag}(\mathbf{r})\Psi_{g m'}(\mathbf{r})\Psi_{e m}(\mathbf{r})\nonumber \\
&&H_{hy}=g_{l}\sum_{m}\int d^{3}\mathbf{r}(\Psi_{e m}^{\dag}(\mathbf{r})\Psi_{g m}(\mathbf{r})+\Psi_{g m}^{\dag}(\mathbf{r})\Psi_{e m}(\mathbf{r})),\label{eq1}
\end{eqnarray}
where fermion field operator $\Psi_{\alpha m}^{\dag}(\mathbf{r})$ creates an atom at position $\mathbf{r}$ with internal state $|\alpha m\rangle$, $\rho_{\alpha m}=\Psi_{\alpha m}^{\dag}(\mathbf{r})\Psi_{\alpha m}(\mathbf{r})$ and $\rho_{\alpha}=\sum_{m=-I}^{I}\rho_{\alpha m}$.
Here, $\alpha=g$ or $e$ denotes electronic state $^{1}S_{0}$ (ground state) and $^{3}P_{0}$ (excited state), respectively. Furthermore, the nuclear Zeeman states of alkaline-earth metal atoms are represented by $m=-I, -I+1,..., I-1, I$. For example, $^{87}$Sr has $I=9/2$, $^{171}$Yb has $I=1/2$ and $^{173}$Yb has $I=5/2$.
The external potential $V_{\alpha}(\mathbf{r})$ is provided by the optical lattice and both periodic potential and harmonic trapping potential are included.
The energy shift of excited state to ground state is $\hbar\omega_{0}$ and $g_{gg}, g_{ee}, g_{eg}^{+}, g_{eg}^{-}$ are all interaction parameters. However, 
because the alkaline-earth metal atoms have a fully occupied outer shell and their total electron spin is zero, the conventional magnetic Feshbach resonance
does not work to tune the interaction between two atoms. Fortunately, the newly proposed orbital Feshbach resonance is able to tune the interaction in these alkaline-earth atom system and has been realized for $^{173}$Yb atom gas\cite{ZhangRen2015,Pagano2015,Riegger2015}.

We should point out that $H_{G}$ has been introduced by Gorshkov et al. to describe a scheme for simulation of Kondo lattice model\cite{Gorshkov2010},
while the extra Hamiltonian $H_{hy}$ is added in the present work but is not covered in their original proposal.
This extra term with coupling constant $g_{l}$ expresses a transition between excited and ground state (e.g. $^{1}S_{0}$$\rightleftharpoons$$^{3}P_{0}$) of alkaline-earth metal atoms like $^{87}$Sr and $^{173}$Yb, which can be induced by extra laser field\cite{Boyd2007,Campbell2009}. Here, generically, the coupling of different nuclear Zeeman sub-level as $\sum_{m,m'}g_{l}^{m,m'}\int d^{3}\mathbf{r}(\Psi_{e m}^{\dag}(\mathbf{r})\Psi_{g m'}(\mathbf{r})+\Psi_{g m'}^{\dag}(\mathbf{r})\Psi_{e m}(\mathbf{r}))$ is possible but we choose the simplest case $g_{l}^{m,m'}=g_{l}\delta_{m,m'}$ with suitable driving laser frequency and polarization.
In the following discussion, we can see that such extra term will lead to the realization of periodic Anderson model since effective $c-f$ hybridization is induced by this direct transition.
Before leaving this issue, we note that if nearest-neighbor $c-f$ hybridization with spin-dependence can be realized, one expects the topological periodic Anderson model is able to be explored\cite{Dzero2010}, and the physics of the advocated topological Kondo insulator may be inspected from this alternative point of view apart from real-life material SmB$_{6}$.

Then, assuming only the lowest band is involved in our model and we expand
\begin{eqnarray}
\Psi_{g m}(\mathbf{r}) &=& \sum_{j}w_{g}(\mathbf{r}-\mathbf{r}_{j})c_{jm} \\ \label{eq2}
\Psi_{e m}(\mathbf{r}) &=& \sum_{j}w_{e}(\mathbf{r}-\mathbf{r}_{j})f_{jm}, \label{eq3}
\end{eqnarray}
where real Wannier function $w_{\alpha}(\mathbf{r})$ is utilized and
$c_{jm}$ ($f_{jm}$) destroys an atom located site $j$ with internal state $|g m\rangle$ ($|e m\rangle$) .
Inserting these equations into Eq.~(\ref{eq1}) leads to lattice Hamiltonian as follows
\begin{eqnarray}
H&=&-t_{c}\sum_{\langle ij \rangle m}c_{im}^{\dag}c_{jm}-t_{f}\sum_{\langle ij \rangle m}f_{im}^{\dag}f_{jm}\nonumber \\
&&-\mu_{c}\sum_{j m}c_{jm}^{\dag}c_{jm}-\mu_{f}\sum_{j m}f_{jm}^{\dag}f_{jm}\nonumber \\
&&+U_{c}\sum_{j,m<m'}c^{\dag}_{jm}c_{jm}c^{\dag}_{jm'}c_{jm'}\nonumber \\
&&+U_{f}\sum_{j,m<m'}f^{\dag}_{jm}f_{jm}f^{\dag}_{jm'}f_{jm'}\nonumber \\
&&+U_{cf}\sum_{j,m,m'}c^{\dag}_{jm}c_{jm}f^{\dag}_{jm'}f_{jm'}\nonumber \\
&&+U_{ex}\sum_{j,m,m'}c^{\dag}_{jm}f_{jm'}^{\dag}c_{jm'}f_{jm}\nonumber \\
&&+V\sum_{j,m}(c^{\dag}_{jm}f_{jm}+f^{\dag}_{jm}c_{jm}). \label{eq4}
\end{eqnarray}
Here, as often done in simulation of fermionic Hubbard model\cite{Esslinger2010}, only nearest-neighbor hopping is considered and the harmonic trapping potential is not explicitly considered but will be added later.
Furthermore, the single center approximation, which means only interaction at the same site is considered $U_{ijkl}^{\phi}=U_{\phi}\delta_{i=j=k=l}$ with $\phi=c,f,cf,ex$.

The parameters in the lattice model are related to original Hamiltonian via
\begin{eqnarray}
  t_{c}&=&-\int d^{3}\mathbf{r} w_{g}(\mathbf{r})(-\frac{\hbar^{2}}{2M}\nabla^{2}+V_{g}(\mathbf{r}))w_{g}(\mathbf{r}+\vec{\delta})\nonumber \\
  t_{f}&=&-\int d^{3}\mathbf{r} w_{e}(\mathbf{r})(-\frac{\hbar^{2}}{2M}\nabla^{2}+V_{e}(\mathbf{r}))w_{e}(\mathbf{r}+\vec{\delta})\nonumber \\
  \mu_{c}&=&-\int d^{3}\mathbf{r} w_{g}(\mathbf{r})(-\frac{\hbar^{2}}{2M}\nabla^{2}+V_{g}(\mathbf{r})+\hbar\omega_{0})w_{g}(\mathbf{r})\nonumber \\
  \mu_{f}&=&-\int d^{3}\mathbf{r} w_{e}(\mathbf{r})(-\frac{\hbar^{2}}{2M}\nabla^{2}+V_{e}(\mathbf{r})-\hbar\omega_{0})w_{e}(\mathbf{r})\nonumber \\
    U_{c}&=&g_{gg}\int d^{3}\mathbf{r} w_{g}^{4}(\mathbf{r})\nonumber \\
    U_{f}&=&g_{ee}\int d^{3}\mathbf{r} w_{e}^{4}(\mathbf{r})\nonumber \\
    U_{cf}&=&\frac{g_{eg}^{+}+g_{eg}^{-}}{2}\int d^{3}\mathbf{r} w_{g}^{2}(\mathbf{r})w_{e}^{2}(\mathbf{r})\nonumber \\
    U_{ex}&=&\frac{g_{eg}^{+}-g_{eg}^{-}}{2}\int d^{3}\mathbf{r} w_{g}^{2}(\mathbf{r})w_{e}^{2}(\mathbf{r})\nonumber \\
    V&=&g_{l}\int d^{3}\mathbf{r}w_{g}(\mathbf{r})w_{e}(\mathbf{r}),\label{eq5}
\end{eqnarray}
where nearest-neighbor vector $\vec{\delta}$ is defined.

Now, if we adjust $g_{eg}^{+}=g_{eg}^{-}$ via suitable Feshbach resonance technique, the spin-spin exchange interaction $U_{ex}$, which is responsible for simulation of Kondo lattice model, vanishes in this case.
Next, the local interaction $U_{c}$ and hopping $t_{f}$ can also be tuned to be small and thus neglected. The resulting Hamiltonian reads
\begin{eqnarray}
H&=&-t_{c}\sum_{\langle ij \rangle m}c_{im}^{\dag}c_{jm}-\mu_{c}\sum_{j m}c_{jm}^{\dag}c_{jm}-\mu_{f}\sum_{j m}f_{jm}^{\dag}f_{jm}\nonumber \\
&&+U_{f}\sum_{j,m<m'}f^{\dag}_{jm}f_{jm}f^{\dag}_{jm'}f_{jm'}\nonumber \\
&&+U_{cf}\sum_{j,m,m'}c^{\dag}_{jm}c_{jm}f^{\dag}_{jm'}f_{jm'}\nonumber \\
&&+V\sum_{j,m}(c^{\dag}_{jm}f_{jm}+f^{\dag}_{jm}c_{jm}). \label{eq6}
\end{eqnarray}

After re-scaling hopping and chemical potential parameters as $t_{c}\rightarrow t$, $\mu_{c}\rightarrow\mu$, $-\mu_{f}\rightarrow E_{f}$
and re-explain fermion operator $c_{jm}$ ($f_{jm}$) as annihilation operator for conduction electron (local f-electron) with internal 'spin' state $m$, we arrive at the so-called $SU(N)$
periodic Anderson model with spin degeneracy $N=2I+1$\cite{Coleman2015}:
\begin{eqnarray}
H_{SU(N)}&=&-t\sum_{\langle ij \rangle m}c_{im}^{\dag}c_{jm}-\mu\sum_{j m}c_{jm}^{\dag}c_{jm}+E_{f}\sum_{j m}f_{jm}^{\dag}f_{jm}\nonumber \\
&&+V\sum_{j,m}(c^{\dag}_{jm}f_{jm}+f^{\dag}_{jm}c_{jm}) \nonumber \\
&&+U_{f}\sum_{j,m<m'}f^{\dag}_{jm}f_{jm}f^{\dag}_{jm'}f_{jm'} \nonumber\\
&&+U_{cf}\sum_{j,m,m'}c^{\dag}_{jm}c_{jm}f^{\dag}_{jm'}f_{jm'}. \label{eq7}
\end{eqnarray}
Currently, the above $SU(N)$ model may be realized for cases with $N=2,6,10$, which corresponds to $^{171}$Yb, $^{173}$Yb and $^{87}$Sr atom gases. Although the heavy fermion materials with unfilled f-shell usually have
such large degeneracy, the complex chemical environment in realistic materials obscures the underlying physics.
Moreover, large-$N$ expansion is widely utilized in heavy fermion community but the reliability of this theoretical tool is not clear as compared to experiments due to the mentioned complication\cite{Coleman2015}.
In contrast, the purity and tunability of optical lattice system gives us an unique opportunity to directly detect the large-N physics of PAM.

Furthermore, if we are interested in the case with $N=2$, where only two nuclear Zeeman sub-levels are considered, the more familiar $SU(2)$ version is obtained
\begin{eqnarray}
H_{SU(2)}&=&-t\sum_{\langle ij \rangle \sigma}c_{i\sigma}^{\dag}c_{j\sigma}-\mu\sum_{j \sigma}c_{j\sigma}^{\dag}c_{j\sigma}+E_{f}\sum_{j \sigma}f_{j\sigma}^{\dag}f_{j\sigma}\nonumber \\
&&+V\sum_{j\sigma}(c^{\dag}_{j\sigma}f_{j\sigma}+f^{\dag}_{j\sigma}c_{j\sigma})+U\sum_{j}n_{j\uparrow}^{f}n_{j\downarrow}^{f} \nonumber \\
&&+U_{cf}\sum_{j}n_{j}^{c}n_{j}^{f} \label{eq8}
\end{eqnarray}
with effective electronic spin $\sigma=\uparrow,\downarrow$, $U_{f}\rightarrow U$, $n_{j\sigma}=f_{j\sigma}^{\dag}f_{j\sigma}$ and $n_{j}^{f}=\sum_{\sigma}n_{j\sigma}$.
Finally, a simplification can be made when $U_{cf}$ is small compared to other energy scales and the system is far away from f-electron valence transition, then the standard PAM reads
\begin{eqnarray}
H_{PAM}&=&-t\sum_{\langle ij \rangle \sigma}c_{i\sigma}^{\dag}c_{j\sigma}-\mu\sum_{j \sigma}c_{j\sigma}^{\dag}c_{j\sigma}+E_{f}\sum_{j \sigma}f_{j\sigma}^{\dag}f_{j\sigma}\nonumber \\
&&+V\sum_{j\sigma}(c^{\dag}_{j\sigma}f_{j\sigma}+f^{\dag}_{j\sigma}c_{j\sigma})+U\sum_{j}n_{j\uparrow}^{f}n_{j\downarrow}^{f}.\label{eq9}
\end{eqnarray}
Because the alkaline(rare)-earth metal atom $^{171}$Yb has nuclear spin $I=1/2$, it has potential to realize the $SU(2)$ PAM in terms of this ultracold atom gas.

Until here, we have not specified the lattice structure, which those ultracold atom gases are loaded onto.
Generally, one-dimensional chain, non-frustrated square, cubic and honeycomb lattice, frustrated triangular and Kagome lattice have been realized in optical lattice systems\cite{Bloch2012}.
The discussion below will focus on one-dimensional chain and two-dimensional square lattice to give a physically transparent explanation of underlying physics and other more complex lattices are straightforward to explore but not shown in present work\cite{Zhong2013}.
\section{Basic features at high temperature}\label{sec3}
The discussion of generic properties of PAM Eq.~(\ref{eq7}$-$\ref{eq9}) can be found in Ref.~\onlinecite{Hewson1993,Coleman2015,Tsunetsugu1997} and the interested reader may refer to these references.

In current experiments of ultracold atom in optical lattice, particularly for realized fermionic Hubbard model, the achieved lowest effective temperature
is about $T\sim t$\cite{Hart2015,Greif2016,Cheuk2016}.
Although both the previously discussed Kondo lattice and our proposed periodic Anderson model have not been realized in alkaline-earth metal atoms,
we expect near-future experiments in this direct may reach similar temperature regime $T\sim t$.

Actually, for $T\geq t$, the system works at relatively high temperature regime since the typical low-energy spin interaction scale, namely Kondo exchange coupling
$J_{K}\simeq \frac{V^{2}}{U}$
is usually smaller than $t$ for intermediate or large $U$ ($U\geq 4t$, $U\gg V$). (The heavy fermion compounds have large $U$ compared to its band-width).
This means $T\geq J_{K}$, thus the corresponding Kondo physics ($T_{K}\sim te^{-t/J_{K}}$) or more explicitly the lattice Kondo effect
cannot be captured in our cases.
At the same time, the competitor of Kondo screening, the long-ranged RRRY magnetic interaction ($T_{RKKY}\sim J_{K}^{2}/t<T$),
is still out of the detectable high temperature regime.

Although the mentioned low-temperature physics is beyond our reach, we still can observe many interesting phenomena in the high temperature regime, which is able to be detected by current cold atom techniques.
For realistic cold atom system, the external trapping potential, which is usually approximated as harmonic potential to confine atoms, should be added in the models and may be simply considered as
a site-dependent chemical potential as\cite{Esslinger2010}
\begin{eqnarray}
H_{PAM}&=&-t\sum_{\langle ij \rangle \sigma}c_{i\sigma}^{\dag}c_{j\sigma}+\sum_{j \sigma}\left(\frac{1}{2}\omega^{2}(\mathrm{r}_{j}-\mathrm{r}_{0})^{2}-\mu\right)c_{j\sigma}^{\dag}c_{j\sigma}\nonumber \\
&&+\sum_{j \sigma}\left(\frac{1}{2}\omega^{2}(\mathrm{r}_{j}-\mathrm{r}_{0})^{2}+E_{f}\right)f_{j\sigma}^{\dag}f_{j\sigma}\nonumber \\
&&+V\sum_{j\sigma}(c^{\dag}_{j\sigma}f_{j\sigma}+f^{\dag}_{j\sigma}c_{j\sigma})+U\sum_{j}n_{j\uparrow}^{f}n_{j\downarrow}^{f}.\label{eq10}
\end{eqnarray}
Here, for simplicity, we have used the standard PAM but the more involved $SU(N)$ version is readily to obtain.
The harmonic trapping potential is now represented as $\frac{1}{2}\omega^{2}(\mathrm{r}_{j}-\mathrm{r}_{0})^{2}$ with the central site as $\mathrm{r}_{0}$ and possible anisotropy of this harmonic
potential is neglected since no qualitative changes appear\cite{Greif2016}.

As studied in Hubbard model with harmonic potential, Mott physics without magnetism can be clearly seen in this system since charge excitation dominates when $T\geq t$\cite{Greif2016,Cheuk2016,Leo2008,Scarola2009,Chiesa2011}.
For PAM in trapping potential like Eq.~(\ref{eq10}), based on our deep understanding on uniform cases without external potential\cite{Pepin2007,Zhong2012,Senthil2004,Vojta2010}, we expect orbital-selective Mott transition should emerge where Mott localization develops only for f-electron while conduction electron is intact. If one only detects f-electron, this may be treated as Mott physics for f-electron and in the remaining part of this paper, we use Mott instead of more correct name orbital-selective Mott by keeping their tiny difference in our mind.

\subsection{Density distribution and DQMC}
\begin{figure}
\includegraphics[width=0.7\columnwidth]{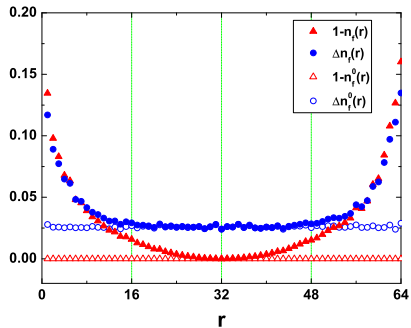}
\caption{\label{fig:1} F-electron real space density $n_{f}(r)$  and variance $\Delta n_{f}(r)$ distribution for $1D$ periodic Anderson model with harmonic potential.
Calculation is performed for a $64$-site system with open boundary condition and setting parameters as $t=1$, $V=1$, $U=12$, $E_{f}=-6$, $\mu=0$, $\omega=0.1$, $\Delta\tau=0.0625$ and $T=1$.
The uniform case with $\omega=0$ is also shown as $n_{f}^{0}(r)$ and $\Delta n_{f}^{0}(r)$}
\end{figure}

\begin{figure}
\includegraphics[width=0.7\columnwidth]{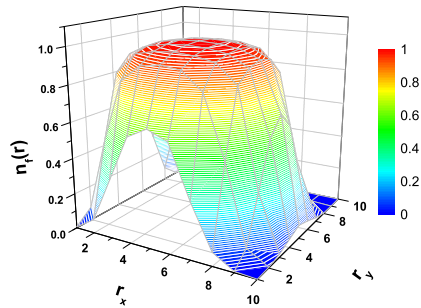}
\caption{\label{fig:2} F-electron real space density distribution for $2D$ periodic Anderson model with harmonic potential.  A $10\times10$ square lattice system with open boundary condition is considered with parameters $t=1$, $V=1$, $U=10$, $E_{f}=-5$, $\mu=0$, $\omega=0.8$, $\Delta\tau=0.125$ and $T=1$.}
\end{figure}

The above expectation is verified by our DQMC calculation in Figs.~\ref{fig:1} and \ref{fig:2},
where occupation of f-electron in real space is plotted and Mott regime with $n_{f}=\langle n_{f}\rangle\simeq1$ is clearly seen at central region while metallic region occupies
the edge part of the system, (statistical errors are very small, at order $0.1$ percent or less for all quantities considered, and not shown) thus coexistence of Mott and metal phase have been observed, which is a generic feature of strongly correlated system in trapping potential\cite{Esslinger2010}.
Meanwhile, in Fig.~\ref{fig:1}, we have also shown the variance of f-electron number at each site $\Delta n_{f}=\langle n_{f}^{2}\rangle-\langle n_{f}\rangle^{2}$ and
its behavior agrees with the results from inspecting occupation number of f-electron $n_{f}$, where charge fluctuation is heavily suppressed at central Mott regime ($\Delta n_{f}\sim0$)
while large variance indicates metallic behaviors around edge parts. It is noted that even in the Mott regime, the variance of f-electron number $\Delta n_{f}$ does not vanish since
only when $U$ is infinite, the charge fluctuation is completely suppressed and gives $\Delta n_{f}=0$. However, for systems with finite $U$ and located in Mott insulating state, charge fluctuation still survives and a small but finite $\Delta n_{f}$ is expected. In our case, we observe $\Delta n_{f}\simeq0.025$ in the central Mott regime.

\begin{figure}
\includegraphics[width=0.4\columnwidth]{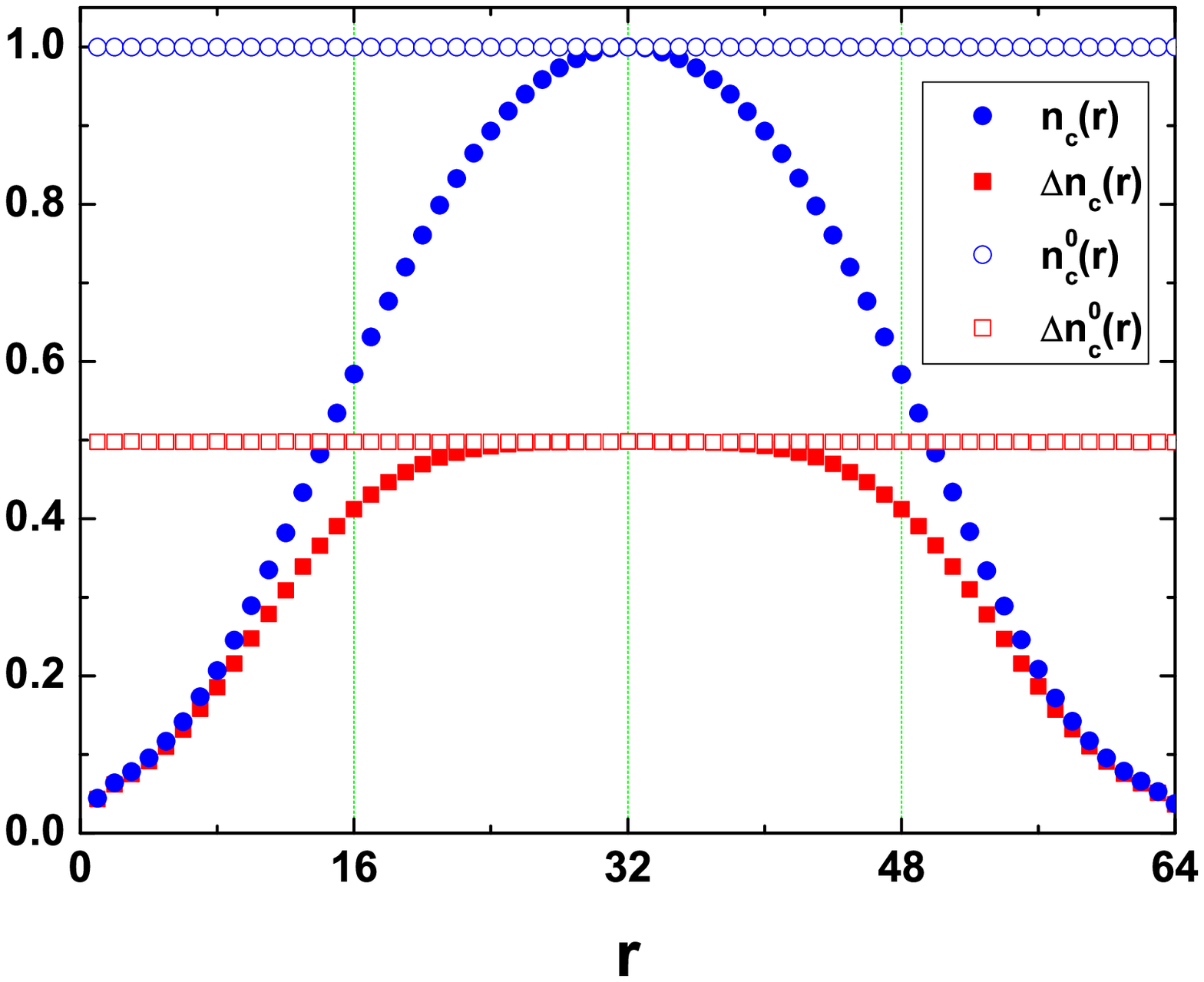}
\includegraphics[width=0.5\columnwidth]{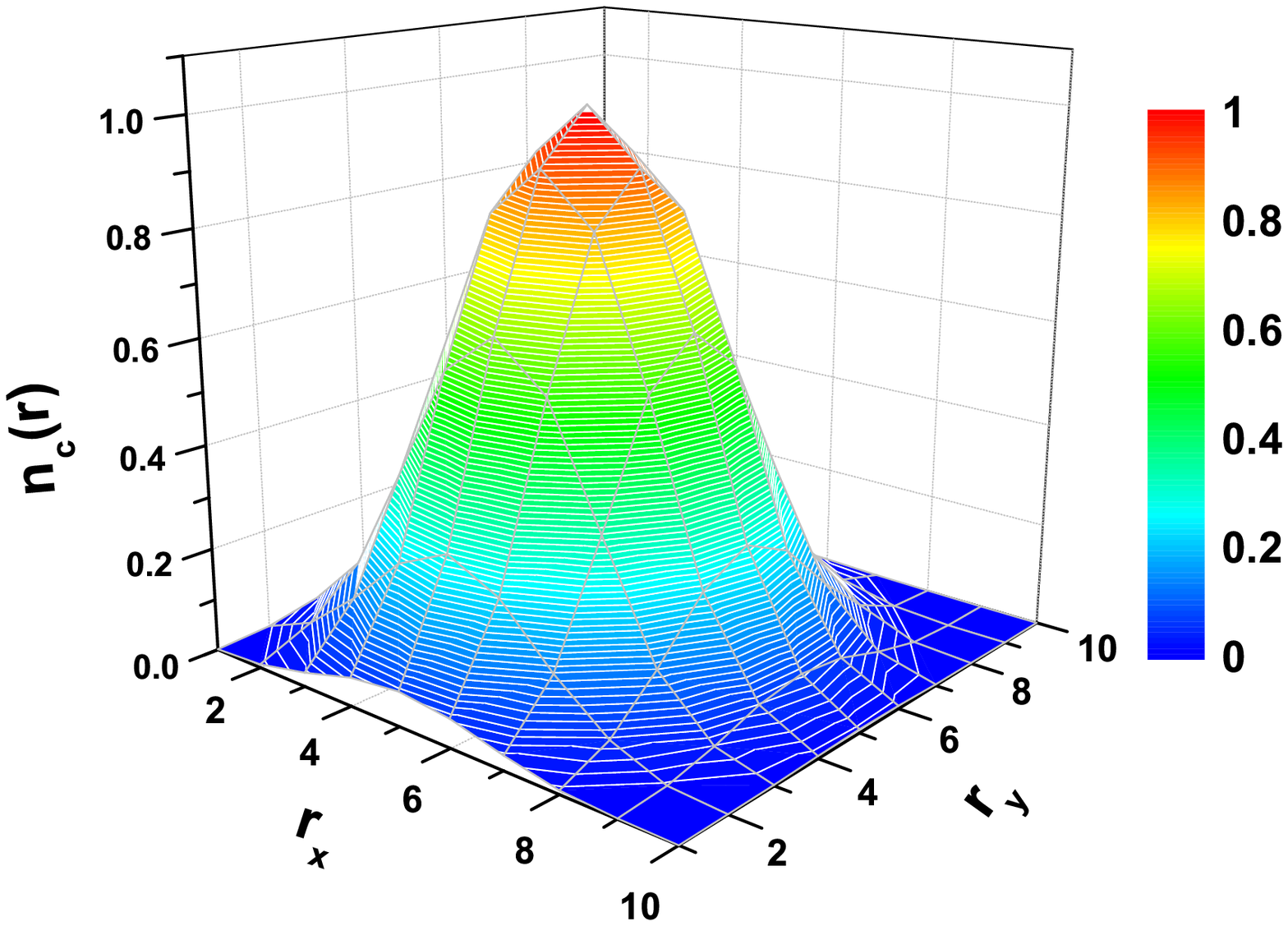}
\caption{\label{fig:2un} Conduction electron density distribution ($n_{c}(r)$, $n_{c}^{0}(r)$) for $1D$ PAM with harmonic potential $\omega=0.1,0$ (left) and $2D$ model with $\omega=0.8$ (right), other parameters are identical to the corresponding cases in Fig.~\ref{fig:1} and Fig.~\ref{fig:2}.}
\end{figure}

For $2D$ system like square lattice as studied in Fig.~\ref{fig:2}, to have a sensible central Mott regime about $20$ sites, a large harmonic potential frequency $\omega=0.8$ is chosen in contrast to the $1D$ chain case with $\omega=0.1$, where most of sites are singly-occupied and show Mott insulating behavior.
Obviously, if harmonic potential is turned off, only Mott state exists in such unform background since the system is half-filling with present parameters,
which can seen by inspecting $n_{f}^{0}$ and $\Delta n_{f}^{0}(r)$, e.g. in Fig.~\ref{fig:1} for a $1D$ case. Additionally, we have checked that small but finite harmonic potential only affect the boundary sites and inner sites show Mott localization as expected. (Density distribution of conduction electron is shown in Fig.~\ref{fig:2un}.)
We note that site-resolved imaging of particle density distribution in fermionic Mott insulator have been recently achieved for $^{6}$Li and $^{40}$K\cite{Greif2016,Cheuk2016},
and our calculated spatial distribution of electrons may be a good guide for realization of PAM in terms of alkaline-earth metal atom gases.

\subsection{Local density of state and DMFT}
Additionally, the Mott behavior of f-electron is able to be seen in their local density of state (DOS) where a large quasi-particle gap (or Mott gap) develops when Mott localization of f-electron is formed.
In experiments, the mentioned Mott gap can be directly probed by measuring the response of the ultracold atom gas in the optical lattice to a modulation of the lattice depth in spite of the intrinsic spatial inhomogeneity
induced by harmonic trapping potential\cite{Jordens2008}. It is noted that this technique has been successfully applied in the simulation of fermionic Hubbard model with $^{40}K$ atoms\cite{Jordens2008}, thus we expect similar techniques may extract information of Mott gap for alkaline-earth metal atoms $^{87}$Sr and $^{173}$Yb in periodic Anderson model regime.

\begin{figure}
\includegraphics[width=0.8\columnwidth]{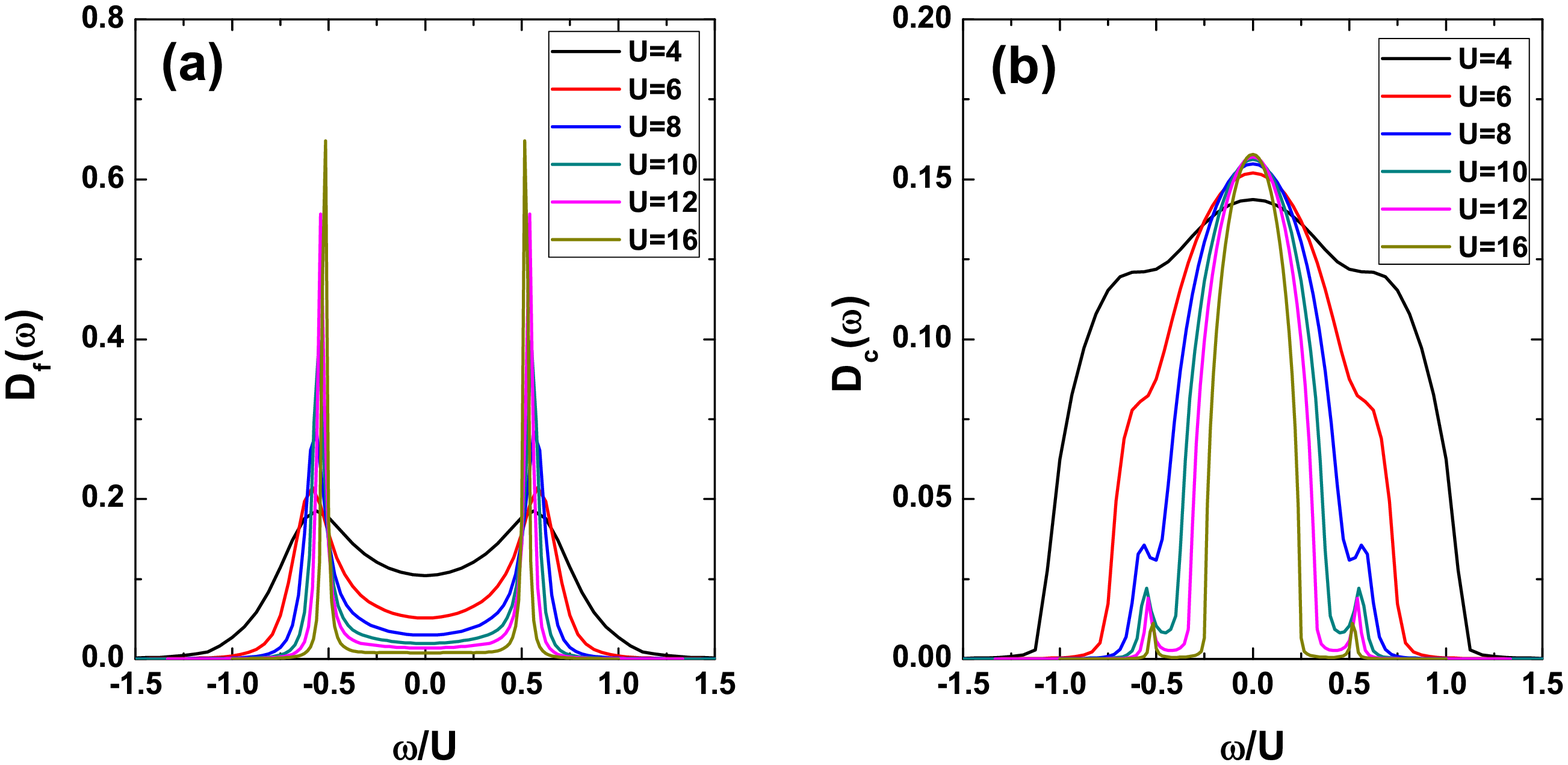}
\caption{\label{fig:3} f-electron density of state $D_{f}(\omega)$ for $2D$ periodic Anderson model on square lattice with parameters $t=1$, $V=1$, $U=4,6,8,10,12,16$, $E_{f}=-U/2$ and $T=1$. For comparison, the
conduction electron density of state $D_{c}(\omega)$ is also shown.}
\end{figure}

Here, due to elusive numerical analytic continuation by transforming imaginary-time quantum Monte Carlo date into real frequency and severe fermion minus-sign problem for larger system size, we do not use DQMC to obtain f-electron DOS but instead, the dynamic mean-field theory (DMFT) approximation is the method of choice. For technical consideration, the harmonic external potential is turned off and the corresponding f-electron DOS for uniform system is shown in Fig.~\ref{fig:3}, which however may be still helpful for understanding on inhomogeneous cases since in experiment (Ref.~\onlinecite{Jordens2008}) performed on inhomogeneous system, one finds similar result as predicted by theory in unform lattice model.
Besides, thermometry, which means to find the effective equilibrium temperature or entropy per particle for experimental samples, is actually done via fitting measured data to theoretical results in uniform model Hamiltonian\cite{Paiva2010,Hart2015}.

In Fig.~\ref{fig:3}, in order to have Mott transition and the related Mott insulating state for f-electron, a half-filled paramagnetic system is considered and a fast solver called iterated perturbation theory is used, which is able to obtain qualitative or semi-quantitative predication compared to more exact but time-consuming solvers at least for half-filled system\cite{Georges1996,Jarrell1993}. From calculation of DMFT, we see that a clear excitation gap establishes between two broad bands (so-called upper and lower Hubbard bands) when $U$ is large ($U\gtrsim12$), which means Mott state for local f-electron is formed via large Hubbard $U$ to prohibit double-occupation on each site. As noted previously, this gap can be measured by corresponding techniques and it also can be used to detect the formation of Hubbard bands.
For comparison, the conduction electron DOS $D_{c}(\omega)$ is also shown and it is clear that no qualitative changes appear for conduction electron during the Mott localization of f-electron\cite{Rozenberg1996}.

\section{Conclusion and discussion}\label{sec4}
In conclusion, we have proposed a possible way to realize both $SU(N)$ periodic Anderson model and its standard version,
where universal heavy fermion physics in condensed matter community is able to be detected and clarified by well-developed optical lattice and cold atom techniques.
For SU(2) PAM, the underlying high temperature but still interesting physics is explored by numerical simulations, i.e. DQMC and DMFT,
where spatial distribution of f-electron density and its local density of state are consistent with the expectation of Mott or orbital-selective Mott physics.
These clear signature of interaction-induced (orbital-selective) Mott localization state for local f-electron may be helpful for corresponding experiments in this direction.
Additionally, the $SU(N)$ physics of PAM has not been explored but we expect novel physics beyond Doniach's classic phase diagram of heavy fermion like charge-density-wave, valence-bond-solid and $\eta$-pairing should emerge as inspired by corresponding large symmetry study in $SU(2N)$ Hubbard-like models\cite{Wu2003,Wu2010,Wang2014,Zhou2016}.

Future investigation on detailed spin/charge or even pairing correlation at site-scale is desirable since similar measurements have been performed in fermionic Hubbard model\cite{Hart2015,Parsons2016,Boll2016}.
In addition, if effective temperature of optical lattice can be further lowered, it will be interesting to study the low-energy/temperature lattice Kondo physics and its corresponding two-fluid behavior\cite{Yang2016},
which is proposed to be core actor for almost all of low-temperature anomalous quantum phenomena.

Another interesting issue is to explore the negative-$U$ periodic Anderson model in these ultracold fermion systems, (such negative-$U$ can be tuned by the orbital Feshbeach resonance technique\cite{ZhangRen2015,Pagano2015,Riegger2015})
because such negative $U$ model is able to be (numerically) solved by DQMC and has been argued as a phenomenological Hamiltonian for heavy fermion superconductivity\cite{Ramos2010}.
Future work in this direction is helpful to uncover the nature of this effective model and will be a good starting point for understanding on unconventional paring mechanism in those strongly
correlated electron systems\cite{Scalapino2012,Anderson2004,Lee2006,Zhong2017}.

\begin{acknowledgments}
We thank Congjun Wu for helpful discussion on $SU(N)$ physics and Ren Zhang on orbital Feshbach Resonance.
This research was supported in part by NSFC under Grant No.~$11325417$, No.~$11674139$ and No.~$11504061$, the China Postdoctoral Science Foundation and the Foundation of LCP.
\end{acknowledgments}

\appendix

\end{document}